\documentstyle[epsf,seceq,preprint]{ptptex}

\newcommand{\ol}{\overline}

\def\diag{\mathop{\rm diag}\nolimits}

\preprintnumber[3cm]{
YITP-98-36\\{\tt hep-th/9806162}\\June 1998}

\title{
Baryon Mass and Phase Transitions\\ in Large $N$ Gauge Theory
}

\author{
Yosuke {\sc Imamura}\footnote{{\tt imamura@yukawa.kyoto-u.ac.jp}}
{}\footnote{
Supported in part by the Grant-in-Aid for Scientific
Research from the Ministry of Education, Science, Sports and Culture
(\#9110).}
}

\inst{
Yukawa Institute for Theoretical Physics,\\
Kyoto University, Sakyo-ku, Kyoto 606-8502, Japan
}

\recdate{
\today
}

\abst{
We calculate the baryon mass in ${\cal N}=4$ large $N$ gauge theory
by means of AdS/CFT correspondence and show that it is a truly bound state,
at least in some situations.
We find that a phase transition occurs at a critical temperature.
Furthermore, we find there are bound states of W-bosons in the Higgs phase,
where the gauge group is broken to $SU(N_1)\times SU(N_2)$.
}

\begin{document}

\maketitle

\section{Introduction}
String theory is a convenient tool for analyzing non-perturbative properties
of Yang-Mills theories.
In the last year, Maldacena presented a new approach to an investigation
of large $N$ conformal field theories.\cite{Maldacena}
His idea relies on the fact that D3-branes with large R-R charge $N$
are approximated well by a classical black brane solution of supergravity.
Its near horizon geometry is described by $AdS_5\times M_5$, where
$AdS_5$ is a five-dimensional anti-de Sitter manifold
and $M_5$ is some five dimensional compact space,
which is assumed to be $S^5$ in this paper.
On the supergravity side, the Yang-Mills gauge field and its superpartners
live on the boundary at infinity of the $AdS_5$.
These fields couple to supergravity fields in bulk,
and correlation functions of operators in
the Yang-Mills theory are obtained as classical correlation functions
of bulk fields on $AdS_5\times M_5$.\cite{GKB,Holography}
Quarks and monopoles are represented by fundamental and D-strings, respectively.
By calculating the area of the string world-sheet,
we obtain a quark-quark potential, monopole-monopole potential,
and quark-monopole potential, etc.\cite{WilsonLoop,Q-M,MacroString}
Furthermore, it has been shown that
baryons correspond to D5-branes wrapped around the compact manifold $M_5$.%
\cite{GrossOoguri,WittenBaryon}

The purpose of this paper is to study baryon configurations.
We calculate the baryon mass at zero and finite temperature.
At low temperature, we find that $N$ quarks combine into a truly bound state,
a baryon.
At a critical temperature $T=T_c$, a phase transition occurs,
and in the high temperature phase, baryons decay into free quarks.
In this paper, we refer to massive W-bosons in the Higgs phase
with unbroken gauge group
$SU(N)\times U(1)$ as quarks, because
if the vev of the Higgs field become infinite,
the W-bosons become external quarks.
Additionally, we discuss the Higgs phase
in which gauge symmetry is broken to $SU(N_1)\times SU(N_2)$,
and it is shown that bound states of W-bosons exist.

In this paper, we consider a maximally supersymmetric (${\cal N}=4$) theory.
Therefore, the existence of truly bound states seems strange.
About this contradiction, we give an argument in the Conclusion.

\section{Zero temperature}
First, we consider the ${\cal N}=4$ $SU(N)$ gauge theory at zero temperature.
It is realized as a field theory on $N$ overlapping extremal D3-branes.
If $N$ is sufficiently large, the three-branes
are approximated well by a classical black three-brane solution
of supergravity.
The metric of the solution of three-branes spreading along $x_{0,\ldots,3}$ is
\begin{equation}
ds^2=H^{-1/2}({\bf x})\sum_{\mu=0}^3dx_\mu^2+H^{1/2}({\bf x})d{\bf x}^2,\quad
H({\bf x})=1+\frac{r_0^4}{r^4},
\label{D3metric}
\end{equation}
where ${\bf x}=(x_4,x_5,\ldots,x_9)$ and $r=|{\bf x}|$.
A radius of the horizon $r_0$ is connected with
the R-R charge $N\in{\bf Z}$ of the D3-branes by the relation
\begin{equation}
r_0^4=\frac{2\kappa^2}{4c_5}NT_{\rm D3}=4\pi l_s^4g_{\rm str}N=2\eta l_s^4,
\label{r0is}
\end{equation}
where $T_{D3}=1/(2\pi)^3l_s^4g_{\rm str}$ is the tension
of a single extremal D3-brane,
$c_5=\pi^3$ is the volume of a five-dimensional unit sphere,
$2\kappa^2=(2\pi)^7l_s^8g_{\rm str}^2$ is Newton's constant,
and $\eta=2\pi g_{\rm str}N$ is an effective coupling constant
of the large $N$ gauge theory.
We adopt the convention
in which the string tension is $T_F=1/(2\pi l_s^2)$ and
the string coupling constant $g_{\rm str}$ is
transformed into $1/g_{\rm str}$ by the S-duality transformation.
In the near horizon region $0\leq r\ll r_0$, Eq.(\ref{D3metric})
is reduced to the $AdS_5\times S^5$ metric:
\begin{equation}
ds^2=\frac{r^2}{r_0^2}\sum_{\mu=0}^3dx_\mu^2
    +\frac{r_0^2}{r^2}dr^2+r_0^2d\Omega_5^2.
\label{AdSmetric}
\end{equation}

Because all fields in ${\cal N}=4$ theories belong to the adjoint representation,
we should introduce quarks
belonging to the fundamental representation $N$ of $SU(N)$
 as `external' ones.
On the supergravity side,
quarks are expressed by semi-infinite open strings whose points at one end
are attached on the horizon $r=0$ of the black brane solution.
These external quarks have infinite masses
because of the infinite length of the strings.
We adopt a definition of the string orientation such that quarks are strings
go outward from the horizon.
Antiquarks in the $\ol N$ representation are represented
by strings with the opposite orientation.

As is shown in \citen{WittenBaryon} and \citen{GrossOoguri}, 
junctions of $N$ strings with the same orientation
can be constructed in $AdS_5\times S^5$,
and they are identified with D5-branes wrapped around $S^5$.
The charge of the string endpoints, which couples to the $U(1)$ field
on the compact D5-brane, is canceled by
the Chern-Simons coupling to the R-R five-form field strength
wrapped around $S^5$.
In terms of the field theory,
these configurations are regarded as baryons, bound states of $N$ quarks,
or antibaryons, bound states of $N$ antiquarks.
We identify baryons with wrapped D5-branes
from which $N$ strings stretch outward.
Antibaryons are D5-branes wrapped around $S^5$ in the opposite direction
to that of baryons.

If we wish to calculate the binding energy of the baryon configuration,
we should introduce a `cutoff'
to make masses of quarks and baryons finite.
This is realized by introducing another D3-brane,
on which another endpoint of each string attached.
We will call this D3-brane a `probe' in this paper,
In terms of the field theory,
this configuration corresponds to the Higgs branch of $SU(N+1)$ gauge theory,
where the gauge group is broken to $SU(N)\times U(1)$.
The position of the extra D3-brane on $AdS_5\times S^5$ background
represents the vev of the Higgs field.
If we take the limit in which the Higgs vev becomes infinite, the $U(1)$ factor decouples,
and massive W-bosons belonging to bi-fundamental representations
$(N,-1)$ and $(\ol N,+1)$ become quarks and antiquarks, respectively.
Before taking this limit,
both quarks and baryons have finite masses,
and we can obtain the binding energy
as the difference between the baryon mass and $N$ times the quark mass.
One may expect that by taking the limit after the subtraction
we will get a finite binding energy.
However, this is not the case.
What we are considering is
conformal theory, and the unique scale parameter is the Higgs vev.
Therefore, if we obtain a non-zero binding energy,
it must be proportional to the Higgs vev,
and the limit makes the binding energy infinite.
Therefore, we suppose the Higgs vev is finite.
In this case, although the quarks are none other than massive W-bosons,
we call them just `quarks', because we focus only on the $SU(N)$ factor
of the broken gauge group $SU(N)\times U(1)$.
In field theory, these `quarks' are contained as dynamical objects
as long as the Higgs vev is finite.
However, in what follows, we express them as BPS configurations of open strings
and, probably, this treatment does not respect the dynamics of the quarks.
This may correspond to the quenched approximation, where loops of the quarks are
neglected.

By using background metric (\ref{AdSmetric}), the quark mass is obtained as
\begin{equation}
M_q=T_F\int_0^{r_{\rm probe}}\sqrt{g_{tt}g_{rr}}dr
=\frac{r_{\rm probe}}{2\pi l_s^2},
\end{equation}
where $r_{\rm probe}$ represents the position of the probe.
The mass of the wrapped D5-brane is a product of the D5-brane tension
$T_{D5}=1/[(2\pi)^5l_s^6g_{\rm str}]$, the area of $S^5$,
and `gravitational potential' $\sqrt{g_{tt}}$:
\begin{equation}
M_{\rm D5}(r)
=\sqrt{g_{tt}}\times T_{\rm D5}\times c_5 r_0^5\
=\frac{r}{2\pi l_s^2}\frac{N}{4}.
\label{Bmass}
\end{equation}
In this expression, the variable $r$ represents the position of wrapped D5-brane.
The $r$ dependence of the mass (\ref{Bmass}) implies that
the wrapped D5-brane feels a force $-dM_{\rm D5}/dr$.
This is the gravitational force due to the black three brane at the center $r=0$.
In addition to it, the D5-brane is pulled by $N$ strings attached to it.
Let us assume $k$ strings are stretched between the wrapped D5-brane
and the probe D3-brane and $N-k$ strings between the D5-brane and
the horizon.
Then, the total force that the D5-brane feels is
\begin{equation}
F=-\frac{dM_{\rm D5}(r)}{dr}+kT_F-(N-k)T_F=T_F\left(2k-\frac{5}{4}N\right).
\end{equation}
(We assume the D5-brane is between the horizon and the probe.)
Therefore, if the condition
\begin{equation}
\frac{5}{8}N\leq k
\label{condition}
\end{equation}
holds, the D5-brane is pulled to the probe and becomes stable on it.
On the other hand, if $k$ is smaller than $(5/8)N$,
the D5-brane moves to the horizon and is absorbed into it.
Consequently, we should identify baryons with wrapped D5-branes
at the position of the probe,
and their mass is given by $M_B\equiv M_{\rm D5}(r_{\rm probe})$.

\begin{figure}[hbt]
\centerline{\epsfbox{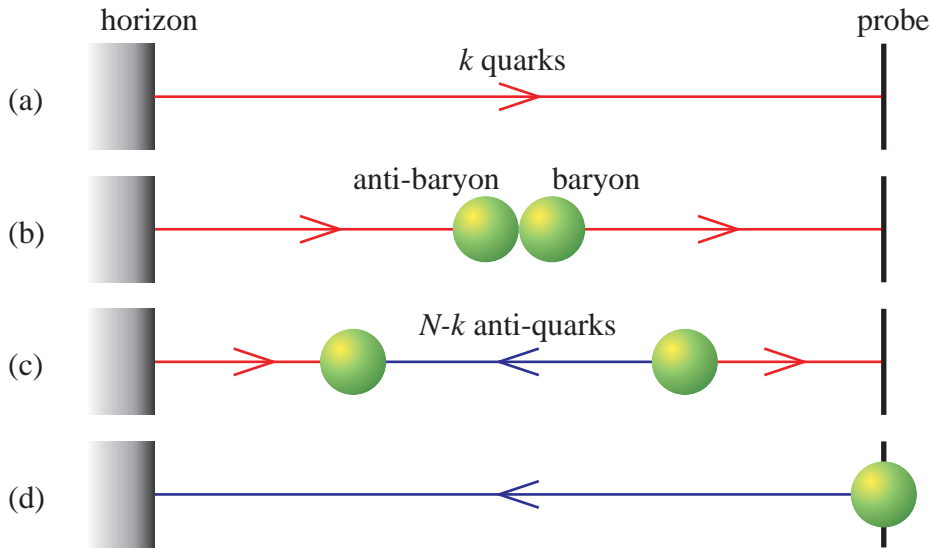}}
\caption{Baryon creation process.
(a) If $k$ strings meet,
(b) two wrapped D5-branes with opposite orientations
are created at a point on
the strings. (c) If the number $k$ of the strings is smaller than $N$,
$N-k$ strings with opposite orientation are generated
between the two wrapped D5-branes
to cancel the charge of string endpoints on the D5-branes.
(d) Then, one of the wrapped D5-branes reaches
to the position of the probe, $r=r_{\rm probe}$,
and another is absorbed into the horizon.}
\label{pair.eps}
\end{figure}

According to the above arguments, it is concluded that
if $N$ quarks meet,
they are bound into a baryon.
Due to positive binding energy,
even if the number of quarks is smaller than $N$, if the cost of
$N-k$ quark-antiquark pair creations is smaller than the baryon binding energy
(This condition is the same as (\ref{condition})),
a baryon is generated.
In terms of the brane configurations, the baryon creation
process may advance as follows (Fig.\ref{pair.eps}):
(a) If $k$ strings meet,
(b) two wrapped D5-branes with opposite orientations
are created at a point on
the strings. (c) If the number $k$ of the strings is smaller than $N$,
$N-k$ strings with opposite orientation are generated
between the two wrapped D5-branes
to cancel the charge of string endpoints on the D5-branes.
(d) Then, one of the wrapped D5-branes reaches
to the position of the probe, $r=r_{\rm probe}$,
and another is absorbed into the horizon.
The final state consists of a wrapped D5-brane
at $r=r_{\rm probe}$ and $N-k$ antiquarks, and the total energy is
\begin{equation}
E_{\rm Final}=(N-k)M_q+M_B=\left(\frac{5}{4}N-k\right)M_q.
\end{equation}
If the condition (\ref{condition}) is satisfied, $E_{\rm Final}$ is smaller
than initial total energy $kM_q$.

\section{Higgs phase}

The Higgs phase where the gauge group $SU(N)$ is broken to a direct product of
its subgroup, is represented by a multi-center three-brane
solution.\cite{Maldacena,HiggsPhase}
In this section we consider the case that
the unbroken gauge group is $SU(N_1)\times SU(N_2)$.
Generalization to an arbitrary number of factor groups is straightforward.
The classical solution is obtained
by replacing the harmonic function $H({\bf x})$
in (\ref{D3metric}) by
\begin{equation}
H({\bf x})=1+\frac{r_1^4}{|{\bf x}-{\bf x}_1|^4}
            +\frac{r_2^4}{|{\bf x}-{\bf x}_2|^4},
\end{equation}
where ${\bf x}_1$ and ${\bf x}_2$ are the positions of two D3-branes,
and the radii $r_i$ are given by
\begin{equation}
r_i^4=4\pi l_s^4g_{\rm str}N_i, \quad i=1,2.
\end{equation}
The near horizon geometry of this classical solution
consists of three parts:
\begin{itemize}
\item region 1 ($|{\bf x}-{\bf x}_1|\ll\frac{r_1}{r_1+r_2}|{\bf x}_1-{\bf x}_2|$),
\item region 2 ($|{\bf x}-{\bf x}_2|\ll\frac{r_2}{r_1+r_2}|{\bf x}_1-{\bf x}_2|$),
\item region 3 ($|{\bf x}-{\bf x}_{1,2}|\gg|{\bf x}_1-{\bf x}_2|$).
\end{itemize}
These three regions are $AdS_5\times S^5$ with radii $r_1$, $r_2$ and $r_0$,
respectively.
\begin{figure}[hbt]
\centerline{\epsfbox{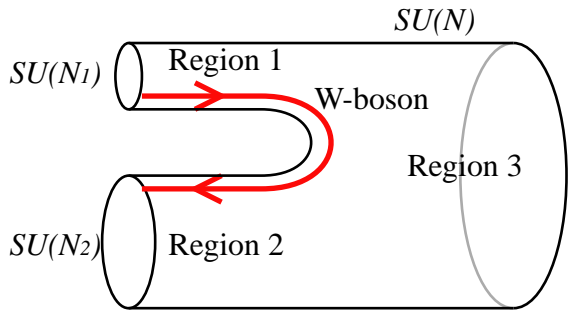}}
\caption{The near horizon geometry of a Higgs phase consist of three regions.
W-bosons which belong to the $(N_1,\ol N_2)$ representation
are represented by strings stretched from the horizon in region 1
to that in region 2.
They are regarded as combinations of $SU(N_1)$ quarks in region 1 and
$SU(N_2)$ antiquarks in region 2.
W-bosons in the $(\ol N_1,N_2)$ representation correspond to strings
with the opposite orientation.
}
\label{higgs.eps}
\end{figure}
\begin{figure}[hbt]
\centerline{\epsfbox{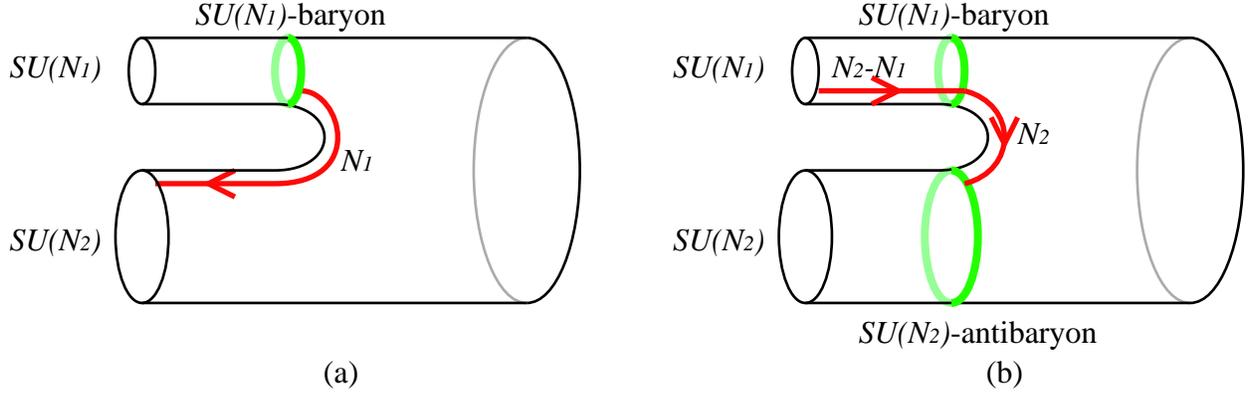}}
\caption{If $k$ W-bosons belonging to the $(N_1,\ol N_2)$ representation
meet, a bound state is created
if $k$ is larger than $(5/8)N_1$ or $(5/8)N_2$.
(a) When $(5/8)N_1\leq k\leq(5/8)N_2$,
it is regarded as combination of an $SU(N_1)$ baryon and $SU(N_2)$ antiquarks.
(b) When $k$ is larger than both $(5/8)N_1$ and $(5/8)N_2$,
the bound state is regarded as a combination of an $SU(N_1)$ baryon
and an $SU(N_2)$ antibaryon.}
\label{combi.eps}
\end{figure}
Though we can discuss quarks and baryons by introducing another D3-brane
probe in the same way as in the last section,
we focus on massive W-bosons here.
Massive W-bosons belonging to the $(N_1,\ol N_2)$ representation
of the unbroken gauge group $SU(N_1)\times SU(N_2)$
are represented by open strings going
from the horizon in region 1 to that in region 2.
This configuration can be regarded
as a combination of a quark configuration in region 1 and
an antiquark configuration in region 2.
W-bosons in the $(\ol N_1,N_2)$ representation are represented by strings with
the opposite orientation.
Therefore, we can discuss the dynamics of the Higgs phase as
in the last section.
In each region, W-bosons behave like quarks or antiquarks.
If their number exceeds the critical value $(5/8)N_1$ or $(5/8)N_2$,
they are bound into an (anti)baryon.
For the sake of a concrete argument,
let us suppose $N_1<(5/8)N_2$ and $N_2<(13/8)N_1$.
If the number $n_W$ of W-bosons in the $(N_1,\ol N_2)$ representation
satisfies $(5/8)N_1\leq n_W<(5/8)N_2$,
an $SU(N_1)$ baryon is created in region 1, while antiquarks in region 2
are not bound.
(a) in Fig.\ref{combi.eps} displays a configuration in the case of $n_W=N_1$.
It consists of an $SU(N_1)$ baryon and $N_1$ $SU(N_2)$ antiquarks.
If the number $n_W$ become larger than $(5/8)N_2$,
an $SU(N_2)$ antibaryon is created in region 2.
(b) in Fig.\ref{combi.eps} is a configuration with $n_W=N_2$.
It consists of an $SU(N_1)$ baryon, an $SU(N_2)$ antibaryon and
$N_2-N_1$ $SU(N_1)$ quarks.
These baryons and antibaryons are confined in regions 1 and 2,
respectively, because of the topology of the spacetime.
Even if $SU(N_1)$ baryons and $SU(N_2)$ antibaryons contact at
the boundary of region 1 and region 2,
they cannot merge into wrapped D5-branes,
because the D5-branes have opposite wrapping directions.

\section{Finite temperature}

Next, we consider baryon configurations at finite temperature.
We again restrict our argument to the unbroken $SU(N)$ gauge theory with a probe
at $r=r_{\rm probe}$.
A gauge theory at finite temperature is realized as a field theory
on non-extremal D3-branes.
The classical non-extremal black three-brane solution of supergravity
has been given by Horowitz and Strominger.\cite{HorowitzStrominger}
In our convention, the Euclidean version of the metric can be written as
\begin{equation}
ds^2=f_+f_-^{-1/2}dt^2+f_+^{-1}f_-^{-1}dr^2+r^2d\Omega_5^2+f_-^{1/2}\sum_{i=1}^3dx_i^2,\quad
f_\pm=1-\frac{r_\pm^4}{r^4},
\label{D3inHS}
\end{equation}
where $r_\pm$ are represented by energy density $E$ and pressure $P$
on the three-brane\footnote{$E$ and $P$ are defined
as the diagonal components of energy-momentum tensor. Namely,
$T_{\mu\nu}=\diag(E,P,P,P)$.} as follows:
\begin{equation}
r_+^4=\frac{2\kappa^2}{4c_5}\frac{3E+P}{4},\quad
r_-^4=\frac{2\kappa^2}{4c_5}\frac{-E+5P}{4}.
\end{equation}
Their geometric mean, $\sqrt{r_+r_-}$, is equal to $r_0$ given by eq.(\ref{r0is}).
If $r_-$ and $r_+$ differ,
this manifold is everywhere smooth and its topology
is $R^3\times D$, where $D$ is a two-dimensional disc
parameterized by $r$ and $t$.
The center of the disc $D$ corresponds to the horizon $r=r_+$.

Because the unique dimensionless parameter of this theory is $M_q/T$,
the massless quark limit is equivalent to the high temperature limit.
In this limit, the probe approaches the center of the disc
$r_{\rm probe}\rightarrow r_+$, and
world sheets of strings which represent quarks become a small disc.
The quark mass is proportional to the area of the small disc.
On the other hand, the baryon mass is proportional to the circumference
of the boundary of the small disc.
Therefore, if the quark mass is smaller than some critical value,
the baryon mass becomes larger than $N$ times the quark mass.
If so, quarks do not make bound states.
In other words, baryons, which are stable at zero temperature,
decay into free quarks at a certain critical temperature $T_c$.

To discuss a near-horizon geometry of the manifold (\ref{D3inHS}),
it is convenient to introduce the new coordinate $\rho$ by
\begin{equation}
\rho^4=\frac{r^4-r_-^4}{r_+^4-r_-^4}.
\label{rhois}
\end{equation}
After proper rescaling of the coordinates $x_{1,2,3}$ and $t$,
we obtain a near-horizon metric:
\begin{equation}
ds^2=r_+^2\left[\left(\rho^2-\frac{1}{\rho^2}\right)dt^2
+\left(\rho^2-\frac{1}{\rho^2}\right)^{-1}d\rho^2
+\rho^2\sum_{i=1}^3dx_i^2+d\Omega_5^2\right].
\label{near-ned3}
\end{equation}
Requiring smoothness at the horizon $\rho=1$,
the period of time $t$ is fixed as follows:
\begin{equation}
0\leq t<\pi.
\end{equation}
On this manifold, the quark mass $M_q$ and the baryon mass $M_B$ are given by
\begin{equation}
\beta M_q
=\frac{1}{2\pi l_s^2}\int_0^\pi dt\int_1^{\rho_{\rm probe}}
                                  d\rho\sqrt{g_{tt}g_{\rho\rho}}
=\frac{1}{2\pi l_s^2}r_+^2\pi(\rho_{\rm probe}-1)
=\sqrt{\frac{\eta}{2}}(\rho_{\rm probe}-1),
\end{equation}
\begin{equation}
\beta M_B
=\frac{1}{(2\pi)^5l_s^6g_{\rm str}}\times \pi^3 r_0^5 \times \pi r_0\sqrt{\rho^2-\frac{1}{\rho^2}}
=\frac{N}{4}\sqrt{\frac{\eta}{2}}\sqrt{\rho^2-\frac{1}{\rho^2}}.
\end{equation}
(The inverse temperature $\beta$ and masses $M_q$ and $M_B$ should be defined
by using the metric in the asymptotic flat region at infinity.
However, their products, $\beta M_q$ and $\beta M_B$, are dimensionless,
and they do not depend upon the metric.)
The ratio of these masses is
\begin{equation}
\frac{M_B}{M_q}=\frac{\sqrt{\rho^2-\frac{1}{\rho^2}}}{4(\rho-1)}N.
\label{ratio}
\end{equation}
In the low temperature limit $\rho\rightarrow\infty$,
this expression reduces to $M_B/M_q=N/4$, which
coincides with the result of the previous section.
\begin{figure}[hbt]
\centerline{\epsfbox{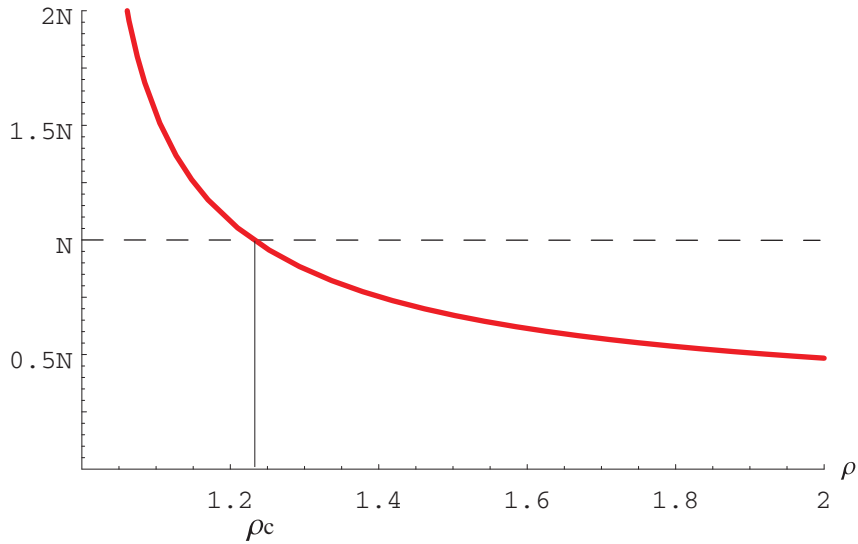}}
\caption{$\rho$-dependence of the ratio $M_B/M_q$.}
\label{plot2.eps}
\end{figure}
The ratio (\ref{ratio}) is a monotonically decreasing function of $\rho$,
and it crosses the line $M_B/M_q=N$ at $\rho\sim1.23$ (Fig.\ref{plot2.eps}).
This means a phase transition occurs at the critical temperature $T_c$
determined by
\begin{equation}
\sqrt{\frac{2}{\eta}}\frac{M_q}{T_c}\sim0.23.
\end{equation}
In the high temperature side $T>T_c$, baryons are unstable
and decay into $N$ quarks.

\section{Conclusion and Discussion}

We calculate the mass of the baryon configurations in the $AdS_5\times S^5$ spacetime
and obtain a value smaller than constituent quark mass.
This implies that quarks are bound into baryons.
The binding energy per quark is of the order of the quark mass,
and even if the number of quarks $k$ is smaller than $N$,
a baryon is created if the binding energy is larger than the cost of
creating $N-k$ quark-antiquark pairs.
In the Higgs phase, where gauge symmetry is broken to $SU(N_1)\times SU(N_2)$,
W-bosons in the $(N_1,\ol N_2)$ representation are regarded as combinations
of $SU(N_1)$ quarks and $SU(N_2)$ antiquarks,
and they also create bound states.
Then we can calculate the baryon mass at finite temperature, and we find that
at a critical temperature, which is the same order as the quark mass,
baryons decay into free quarks.

There is one point that seems strange.
The theory which we have considered has
maximal (${\cal N}=4$) supersymmetries.
Although some of these are broken by means of quarks or baryons,
others may be left unbroken,
and they guarantee the stability of the states.
Therefore, even if bound states are generated,
they should be marginal ones.
This seems to contradict our analysis of baryon configurations.
How can we resolve this contradiction?

One possibility is that our treatment of the brane configuration is improper.
Namely, in this paper,
we neglected the deformation of the wrapped D5-brane
and the electric field on the brane.
Actually, the D5-brane is deformed by means of the tension of strings attached to it.
If the compact manifold is $T^n$ instead of $S^5$,
it is known that the decrease of energy due to the deformation of the D-brane
is exactly canceled by the energy of the electric field on the D-brane
by means of the BPS condition.\cite{BIstring1,BIstring2}
Therefore, in this case, although the deformation of the D-brane
and the energy of the gauge field on the D-brane cannot be neglected separately,
we can treat the configuration
as a combination of strings and a flat D-brane wrapped around $T^n$
for the purpose of calculating the total energy of the brane configuration.
On the other hand, it is not clear whether this treatment can be justified for
the baryon configuration on $AdS_5\times S^5$.

We can construct a configuration without such a problem.
Let us consider a configuration which consist of
a D5-brane wrapped around $S^5$ and
$N$ open strings scattered uniformly over the $S^5$.
In this case, because $N$ string endpoints are distributed
on the D5-brane, if $N$ is sufficiently large,
the D5-brane is almost spherical, and
the electric field on the brane almost vanishes.
(Of cause, in this case, it is necessary to introduce many D3-brane probes.)
Furthermore, all supersymmetries are broken in this configuration,
and a positive binding energy does not contradict the BPS bound.
The BPS bound is expressed by the equation
\begin{equation}
m\geq\left|\sum_iZ_i\right|,
\end{equation}
where $Z_i$ represent the central charges, and the index $i$ labels
the constituents of the system.
In the present case, $i$ labels the $N$ strings, and
the central charges $Z_i$ are vectors in 6-dimensional space.
($Z_i$ belong to $\bf 6$ of $SU(4)$ R-symmetry.)
The direction of the vector corresponds to the position of the open string on $S^5$.
If all the strings lie on one point on $S^5$, the charges $Z_i$ also have the same
direction, and we have
\begin{equation}
\left|\sum_iZ_i\right|=\sum_i\left|Z_i\right|.
\end{equation}
This implies that the bound state is marginal.
On the other hand, if strings are scattered over the $S^5$,
the charges $Z_i$ have different directions, and
the BPS bound is smaller than the sum of the quark masses $|Z_i|$:
\begin{equation}
\left|\sum_iZ_i\right|<\sum_i\left|Z_i\right|.
\end{equation}
Therefore, in this case, the supersymmetries and the existence of truly bound states
do not contradict each other.

\section*{Acknowledgements}
I would like to thank H.Kunitomo
for carefully reading the manuscript.

\bigskip\noindent
{\bf Note Added:} After this work was completed, a paper \citen{BISY} appeared 
which discusses the same issue.

\end{document}